\documentclass[useAMS]{mn2e}
\input{epsf}
 \usepackage[dvips]{graphicx}

\title[The iron line in the low/hard state in GX339--4]{A re--analysis
  of the iron line in the XMM-Newton data from the low/hard state in GX339--4}

\author[C. Done, \& M. Diaz Trigo]
{Chris Done$^{1}$, \& Maria Diaz Trigo$^2$ \\ 
$^1$Department of Physics, Durham University, South Road, Durham, DH1
  3LE, UK: chris.done@durham.ac.uk\\
$^2$ XMM-Newton Science Operations Centre, Science Operations Department, ESAC, PO Box 78, 28691 Villanueva de la Cañada, Madrid, Spain }
\date{}

\def\la{\mathrel{\hbox{\rlap{\hbox{\lower4pt\hbox{$\sim$}}}{\raise2pt\hbox{$<$}}}}}
\def\ga{\mathrel{\hbox{\rlap{\hbox{\lower4pt\hbox{$\sim$}}}{\raise2pt\hbox{$>$}}}}}
\def\ls{\mathrel{\hbox{\rlap{\hbox{\lower4pt\hbox{$\sim$}}}\hbox{$<$}}}}
\def\gs{\mathrel{\hbox{\rlap{\hbox{\lower4pt\hbox{$\sim$}}}\hbox{$>$}}}}

\begin{document}
\maketitle 

\begin{abstract}

The detection of an extremely broad iron line in XMM-Newton MOS data
from the low/hard state of the black hole binary GX339-4 is the only
piece of evidence which unambiguously conflicts with the otherwise
extremely successful truncated disc interpretation of this
state. However, it also conflicts with some aspect of observational
data for all other alternative geometries of the low/hard state,
including jet models, making it very difficult to understand. We
reanalyse these data and show that they are strongly affected by
pileup even with extensive centroid removal as the source is
$\sim$~200$\times$ brighter than the recommended maximum countrate.
Instead, we extract the simultaneous PN timing mode data which should
not be affected by pileup. These show a line which is significantly
narrower than in the MOS data. Thus these data are easily consistent
with a truncated disc, and indeed, strongly support such an
interpretation.

\end{abstract}


\section{Introduction}

The current paradigm for the structure of the accretion flow in black
hole binaries (hereafter BHB) at low luminosities is that the cool,
optically thick, geometrically thin standard accretion disc is
replaced in the inner regions by a hot, optically thin, geometrically
thick flow (Esin et al. 1997). This model has gained widespread
acceptance by its ability to provide a framework in which to interpret
large amounts of apparently unrelated observational data by decreasing
the truncation radius of the disc as the mass accretion rate
increases. At the lowest luminosities, the lack of an inner disc means
that there are few seed photons from the disc illuminating the flow,
so the spectra are hard. Decreasing the disc truncation radius leads
to greater overlap of the flow with the disc, so more seed photons to
Compton cool the flow, giving softer spectra. This also gives a
correlated increase in the amount of reflection and iron line produced
by irradiation of the disc, and an increase in the broadening of these
features by the stronger special and general relativistic effects. The
decreasing radius also means that any frequencies set by this radius
will increase, giving a qualitative (and sometimes quantitative)
description of the increasing characteristic frequencies (including
Quasi- Periodic Oscillations) seen in the power spectra and their
tight correlation with the energy spectra. The flow is completely
replaced by the disc when the disc reaches its minimum radius (last
stable orbit), giving a physical mechanism for the marked hard-to-soft
transition seen in black hole binaries.  Even the jet behaviour can be
tied into this picture, as a large scale height flow is probably
required for jet formation, so the collapse of the inner flow as the
disc reaches its minimum radius triggers a similar collapse of the
radio emission (see e.g. the reviews by Remillard \& McClintock 2006
and Done, Gierlinski \& Kubota 2007, hereafter DGK07, 
with specific data on GX339-4 in e.g. Fender et al 1999; Belloni et al 2005).

Nonetheless, such models are ruled out despite these evident successes
if the disc extends down to the last stable orbit in the low/hard
state. There are two clear claims of this in the literature, first
from a detection of an extremely broad iron line in the low/hard state
of GX339-4 (Miller et al 2006 hereafter M06, Reis et al 2008 hereafter
R08) and secondly from the detection of the residual thermal disc
emission which is consistent with a disc extending down to the last
stable orbit (M06; Rykoff et al 2007; R08, 
Reis, Miller \& Fabian 2009). The direct disc emission can
be alternatively interpreted as a truncated disc, especially when
including the effects of irradiation of the inner disc (Cabanac et al
2009; Gierlinski, Done \& Page 2008). There is also potential
contamination by other components in soft X-rays, as must be the case
in the well studied dim low/hard state spectra from XTE
J1118+480. This system has very low absorption, allowing the cool,
truncated disc emission to be seen in the UV/EUV (Esin et al 2001)
while the soft X-ray excess is clearly hotter and less luminous (Reis,
et al 2009; Chiang et al 2009; perhaps arising from the jet: Brocksopp et al 2010).

Thus the extreme broad iron line in the XMM-Newton data of GX339-4 is
the only strong evidence against the truncated disc/hot flow models
for the low/hard state (Tomsick 2008). If this is a unique
interpretation of the data then it is sufficient evidence to rule out
the hot inner flow/truncated disc geometry and to require an
alternative picture for the low/hard state. However, there are strong
constraints on the geometry of any cool disc with respect to the hard
X--ray source which arise from reprocessing.  We outline these, and
show that the extremely broad line conflicts with some aspect of the
observational data for {\em all} current alternative geometries for
the low/hard state, even that of a mildly relativistic beaming of the
hard X-ray associated with jet model (Beloborodov 1999).

These conflicts motivate us to re-examine these data from the low/hard
state of GX~339-4 to see how robust the iron line parameters are. We
find that while the description of the line in the data is robust, the
data themselves are not. The source is $\sim 200\times$ brighter than
the recommended limit for pile up in the data modes used.  We show
that pileup is still an issue even excluding the central 18" core of
the image and using only single events as in M06. R08 reanalysed these
data excluding a 50" core, using single plus double events. We show
that while the singles are probably free from pileup at this large
radius, the doubles are not, so these data are still affected, plus
there are additional uncertainties form response and background. 

Instead, we extract the PN data which is in timing mode so is not
strongly affected by pile up. These give a line which is clearly
narrower than that of M06 and R08. This narrower line is also
consistent with the simultaneous RXTE data, contrasting with the
extreme broad line from the MOS data which requires the addition of an
'instrumental' edge feature at 4.78~keV in RXTE. We fit both simple and
sophisticated models to the PN spectrum, and show that the line in
these data strongly supports a truncated disc interpretation of the
low/hard state. This conclusion is further strengthened by 
Suzaku data from a lower luminosity low/hard state 
of GX339-4 which also show a narrow line (Tomsick et al 2009).

\section{Potential alternative geometries for the low/hard state}

The main problem is to explain how an extremely broad line can
co-exist with a hard spectrum. The line is produced by X-ray
illumination of the disc and the fraction of emission which is not
reflected is (quasi)thermalised, adding to any intrinsic soft flux
from the disc. Some fraction of this flux is re-intercepted by the
corona, increasing its Compton cooling, and making the spectrum
softer. It is this coupling between the reflected emission and the
corona seed photons via reprocessing which makes it difficult to
maintain a hard spectrum from the corona in the presence of a cool
material which extends down close to the black hole.

For example, if the corona extends smoothly over the disc in a
sandwich geometry then all the reprocessed photons are re-intercepted
by the corona. If the reprocessed emission thermalises, then it adds
to the soft photons for Compton cooling.  Energy balance then gives
spectra which are necessarily steep, in conflict with the observed
hard spectra which define the low/hard state (Haardt \& Maraschi
1993). Compton anisotropy in this plane parallel geometry can harden
the observed spectrum in the first few scattering orders (Haardt \&
Maraschi 1993), but this then predicts a break at higher energies
which is not seen in the data (Barrio, Done \& Nayakshin
2003). Alternatively, extreme ionisation of the disc could prevent
thermalisation of the reprocessed flux, but this requires some very
specific conditions which may not be physical (constant density disc
atmosphere rather than hydrostatic equilibrium: Malzac et al 2005).

Another way to make a hard spectrum with a corona above the disc is if
the corona is patchy rather than smooth, or forms a 'lamppost' above
the spin axis.  In either geometry, a large fraction of the
reprocessed flux need not re-intercept the corona and cool it, so can
produce hard spectra (Stern et al 1995; Malzac et al 2005), while the
illumination of the disc gives the line. However, this illumination
should also give rise to an accompanying continuum reflection
signature, and these reflected photons are then likewise not
re-intercepted, predicting a strong reflection signature to accompany
a strong, broad line. Yet the observed amount of reflected continuum
(and line) is generally rather weak in the low/hard state (around
$\sim 0.3\times$ that which would be expected if the disc covered half
the sky as seen from the X-ray source for these GX339-4 data: see Fig
10 of R08), but both line and amount of reflection increase together
as the spectrum softens (e.g. Gilfanov, Churazov \& Revnivtsev 1999;
Ibragimov et al 2005; DGK07). This can be due to a changing
illuminating geometry or a changing ionisation state (see below).  The
changing geometry has an obvious origin in the truncated disc model,
with increased reflection (and smearing) as the disc extends further
into the hot flow, correlated with a steeper spectral index as more
seed photons illuminate the Comptonising region. Instead, if the disc
inner radius is fixed, then the source geometry has to change.  In a
patchy corona this can be linked to its covering fraction over the
disc, but this gives the opposite trend to that seen in the data.  A
higher covering fraction of the corona over the disc gives steeper
spectra by intercepting more (reprocessed and intrinsic) flux from the
disc but also means that less of the reflected emission can escape
(Malzac et al 2005). Conversely, in a static 'lamppost' model there is
no possible geometry change.

Nonetheless, even without a changing geometry, the contrast of the
reflection features with respect to the illuminating continuum can
change with ionisation state of the material. Extreme ionisation
effects can lead to a suppression of the apparent amount of reflection
below 20~keV as this removes the characteristic atomic features. Thus
the change in amount of reflection discussed above could instead be a
result of changing ionisation rather than geometry. However,
ionisation has no effect on the high energy shape of the reflected
emission as this is determined only by Compton downscattering
losses. Thus the shape of the high energy spectrum (50-200~keV)
constrains the {\em total} amount of reflection, and this clearly
shows that the amount of reflection in the low/hard state is
intrinsically low (Maccarone \& Coppi 2002; Barrio, Done \& Nayakshin
2003).

The combination of observed small solid angle of reflection with a
hard spectrum (i.e. small inferred seed photon flux from reprocessing)
can be reconciled with an underlying disc only if the hard X-rays are
anisotropic. An attractive origin for this is the mildly relativistic
jet seen in the low/hard state, with resultant mild beaming of the
X-ray irradiation away from the disc (Beloborodov 1999, Ferreira et al
2006). These models, with a cool, passive disc down to the last stable
orbit acting as an anchor for the magnetic fields which launch the
jet, then seem to give a possible origin for the broad iron
line. However, the broad line itself is inconsistent with this
geometry as it requires an emissivity which is strongly centrally
peaked (M06; R08). This is opposite to the less centrally concentrated
beaming pattern which arises from a mildly relativistic jet. Instead,
such centrally concentrated emission could be produced either by
strong lightbending for a source very close to the black hole
(Miniutti et al 2003; Miniutti \& Fabian 2004), or by the X-ray source
moving towards the disc e.g.  a failed jet (Henri \& Petrucci 1997;
Ghisellini et al 2004). However, both of these then lead to enhanced
reflection fractions, rather than the small reflection fractions
observed.  The alternative jet models of Markoff, Nowak \& Wilms
(2005) also converge on the truncated disc, hot inner flow geometry,
so do not offer a way to produce the broad line.

Instead, perhaps the combination of scarcity of seed photons to give
the hard spectrum and a (very centrally concentrated) broad iron line
could be reconciled if the cool reflecting material were not a
complete disc, but formed only a small ring close to the innermost
stable orbit. In effect this is a reversal of the truncated disc/hot
flow model, with the disc on the inside, and the hot flow on the
outside. Such a geometry may form close to the major hard-soft
spectral transition from disc evaporation/condensation (Liu et al
2007). While this is a potentially viable model (see also Chiang et al
2009), it runs into subtle problems with the very rapid spectral
variability. The fastest variability has harder spectrum (and less
reflection) than the slower variability (Revnivtsev, Gilfanov \&
Churazov 1999), leading to a complex series of time lags across the
spectrum (Miyamoto \& Kitamoto 1989; Poutanen \& Fabian 1999).  This
can be explained in the standard truncated disc/hot inner flow models
as the larger timescale variability is characteristic of the larger
radii regions of the hot flow. This is the region of overlap between
the hot flow and truncated disc, so giving more seed photons and
reflection. Conversely, the most rapid variability is produced in the
innermost regions, where there is no overlap with the disc and the
spectra are harder (Kotov et al 2001; Arevelo \& Uttley 2006). 
The inner ring of material reverses this, predicting that the
fastest variability has the softest spectra and most reflection,
contrary to the observations. 

Thus while there is an obvious conflict between an extremely
relativistically broadened line and the truncated disc models, all
other alternative geometries currently suggested for the low/hard
state also have issues with this observation. Assuming that the line
shape is robust, and is set by relativistic effects alone (with no
contribution from scattering in an outflow: Titarchuck et al 2009; Sim
et al 2010), then there appears to be no geometry which can
simultaneously explain the line shape together with a hard continuum
spectrum in these data without giving rise to potentially serious
problems with other observations.

\section{XMM-Newton Data Selection and Extraction}

\begin{figure} 
\begin{center} 
\leavevmode  \epsfxsize=8cm \epsfbox{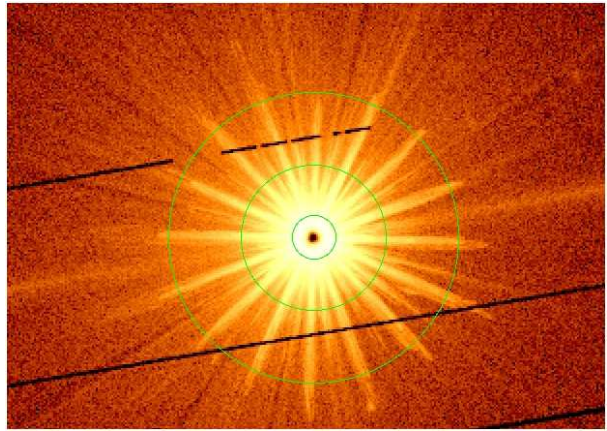} 
\end{center} 
\caption{MOS1 image, with the black centre showing the extreme
  pileup. The circles have radii 18, 60 and 120 arcsec from the
  center. We extract spectra from single events only from 0-120,
  18-120, 60-120 and (not shown) 90-120 arcsecond regions.}
\label{fig:image} 
\end{figure}

\begin{figure*}
\begin{tabular}{cc}
\includegraphics[width=0.45\textwidth,bb=60 40 530 380,
  clip=true]{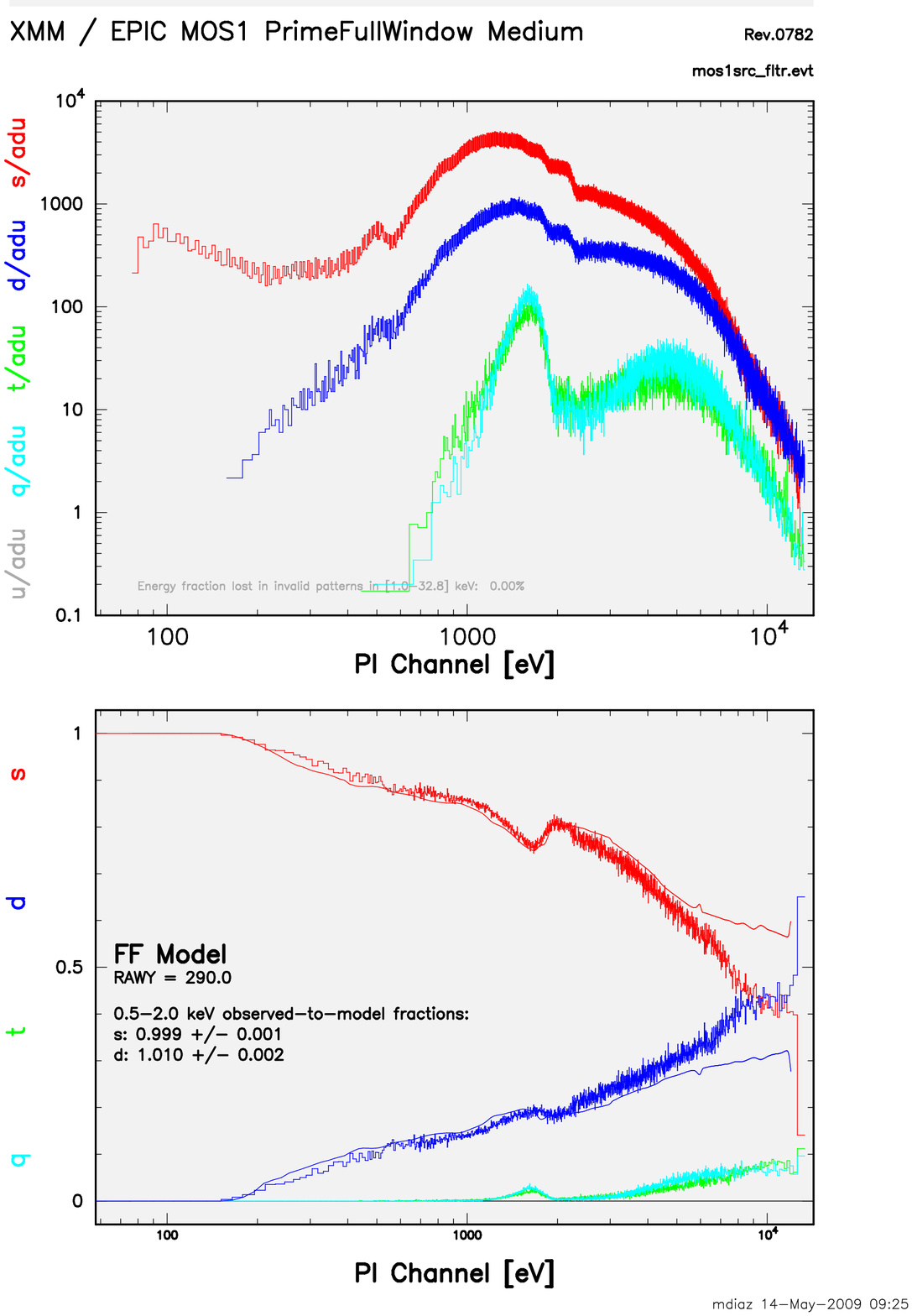}
&
\includegraphics[width=0.45\textwidth,bb=60 40 530 380,
  clip=true]{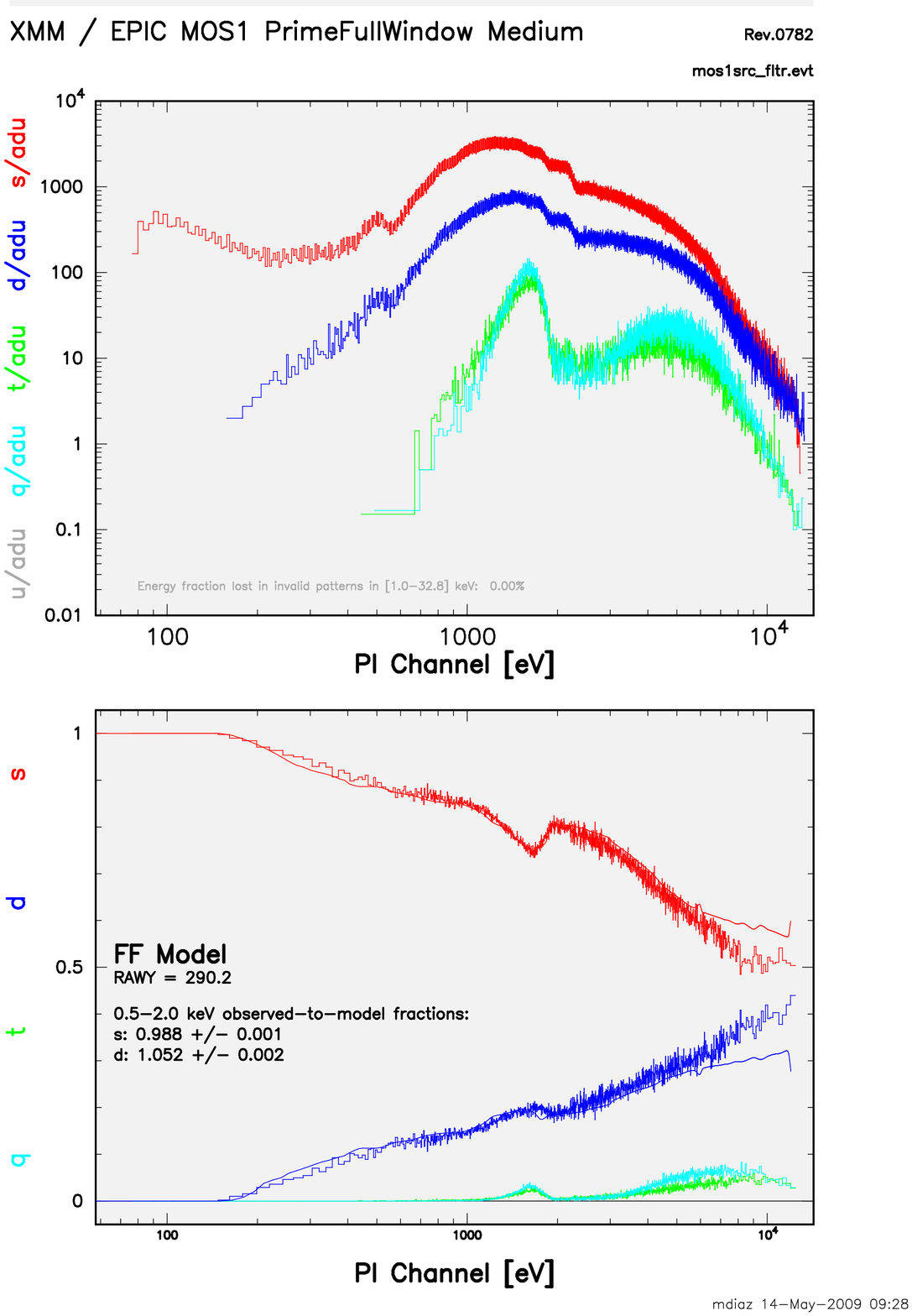}\\

\includegraphics[width=0.45\textwidth,bb=60 40 530 380,
  clip=true]{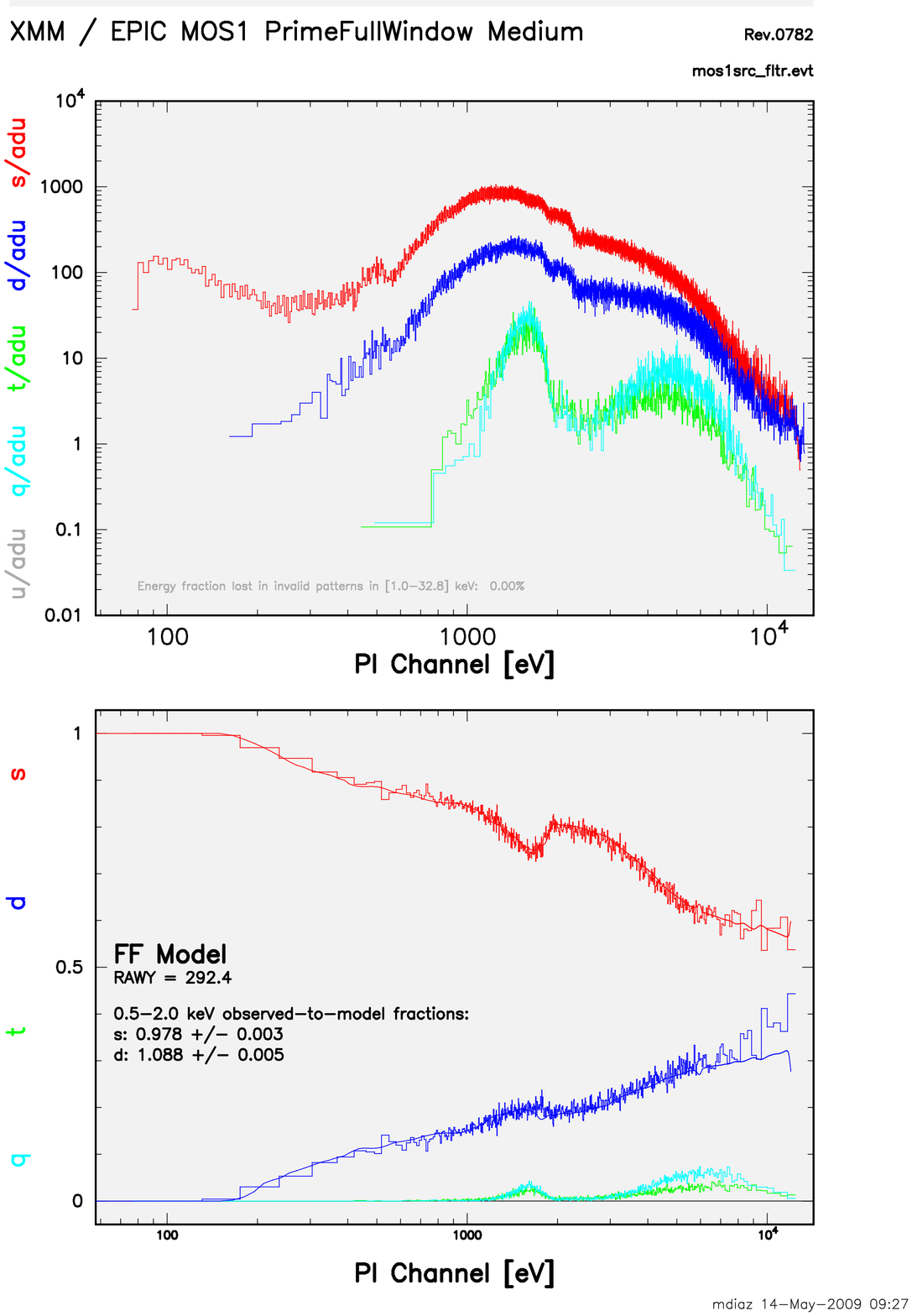}
& 
\includegraphics[width=0.45\textwidth,bb=60 40 530 380,
  clip=true]{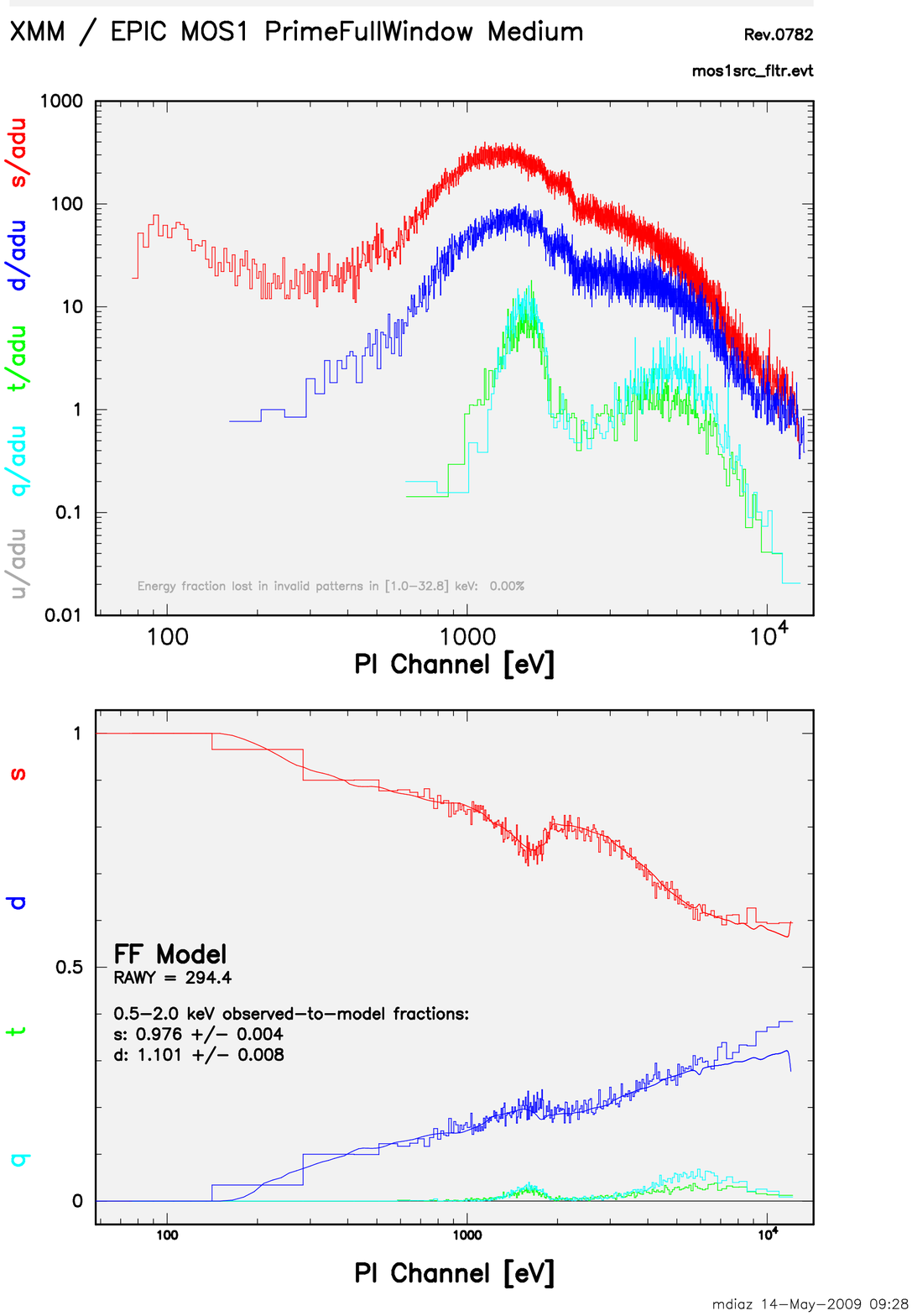}

\end{tabular}

\caption{Quantative measures of pileup in the MOS1 camera using the
SAS task {\tt epatplot}. The data points show the fraction of events
which register as singles (s), doubles (d), triples (t) and quadruples
(q), while the line shows the expected fraction for non-piled up
data. The upper panels show the data extracted from 0-120 and 18-120
arcsec regions, while the lower panels show 60-120 and 90-120
arcseconds. The deviations between the expected and observed pattern
distributions are clearly present even for the singles for the 0-120
arcsec data above 3~keV, and for 18-120 above 6~keV. The 60-180 and
90-120 singles distribution appears clean, though the doubles probably
are still affected above 7~keV.}
\label{fig:epatplot}
\end{figure*}

The XMM-Newton Observatory (Jansen et al 2001) includes three
1500~cm$^2$ X-ray telescopes each with an EPIC (0.1--15~keV) at the
focus. Two of the EPIC imaging spectrometers use MOS CCDs (Turner et
al 2001) and one uses PN CCDs (Struder et al 2001). 

We analysed XMM-Newton observations 0204730201 and 0204730301, taken
in revolutions 782 and 783, starting at 2004-03-16 16:23:41 and
2004-03-18 17:19:49, respectively.  The MOS data are taken in Full
Frame mode, with the medium filter in place, while the PN data are in
timing mode, again with the medium filter. Data products were reduced
using the Science Analysis Software (SAS) version 9.0 (beta).  This
includes an update in calculating the ancillary
response files for non-contiguous extraction regions for timing data, 
as required for excluding the central columns to check for pileup. 

The events list were screened against times with high instrumental
flaring. This resulted in net exposures time of 66, 81 and 81 ks for
EPIC pn, MOS1 and MOS2 cameras for observation 0204730201 and of 53,
70 and 70 ks for EPIC pn, MOS1 and MOS2 cameras for observation
0204730301.

The source was very bright, and the PN data (see below) give a
0.4-10~keV flux of $1.3\times 10^{-9}$~ergs~cm$^{-2}$~s$^{-1}$ for a single
power law fit with $\Gamma=1.55$ and $N_h=4.1\times 10^{21}$
cm$^{-2}$. This gives a WebPIMMS predicted count rate for the MOS in
full frame mode with the medium filter of 61 c/s for a 15'' extraction
region, or total of $\sim 120$~c/s, nearly 200$\times$ larger than the
0.7~c/s recommended total count rate limit for pileup not to affect
the spectra.  Instead, the measured count rate for a 15'' extraction
region is 4.4 c/s, as the central part of the image is so badly piled
up as to be black (see Fig.\ref{fig:image}).

'Photon pile-up' occurs when a source is so bright that there is the
non-negligible possibility that two or more X-ray photons deposit
charge packets in a single pixel during one read-out cycle (i.e. one
frame). This leads to complete flux loss in the central part of the
source if the summed energy of the two photons is larger than the 
reconstructed energy rejection threshold, or spectral hardening 
if the summed energy is below the rejection threshold.
In addition, pattern pile-up occurs when two or more X-ray photons
deposit charge packets in neighbouring pixels within the same frame
e.g. two single soft events will be interpreted as a double event with
twice the energy. Again, this leads to flux loss or spectral hardening
depending on whether the summed energy of the event is above or below
the rejection threshold. 

\subsection{MOS data}

The EPIC MOS Full Frame Mode reads out all pixels of all CCDs,
covering the full field of view.  We first extract a source spectrum
from a 120~arcsec radius circle centred on the source position
(largest circle in Fig. \ref{fig:image}) and illustrate the extent of
pileup on these using the SAS task {\tt epatplot}.
Fig.\ref{fig:epatplot}a shows the distribution of observed single- and
double- pixel (and more complex) events together with their expected
values. The effects of pile-up are evident when the data and model
distributions diverge, as is clearly the case here above 4~keV. The
effects of pile-up will be different for each source, depending on the
spectral shape (see XMM-Newton Users Handbook for some examples).

Pileup cannot yet be corrected, but its effects can be mitigated in
imaging modes by excising the heavily piled-up core from the image,
and extracting counts in an annulus which includes only the lower
count rate wings of the point spread function (see XMM-Newton Users
Handbook). We follow M06 and exclude the inner 18" region (smallest
circle in Fig. \ref{fig:image}). M06 chose this radial range by using
{\tt epatplot} to assess the extent of pileup. We show the {\tt
epatplot} results in Fig.~\ref{fig:epatplot}b. There is still a marked
deviation in the distribution of observed events from the expected
curve above 6~keV (somewhat better than the 4~keV for the full data,
but still clearly affected by pileup). 

{\sc epatplot} also integrates the observed ratios over a given energy
range to quantify the amount of pileup. The default range is 0.5-2~keV
as this is most sensitive to pileup for soft photon spectra. The
observed distribution of single events in this bandpass is within
$\sim 3\sigma$ of the expected value (M06).  However, the spectrum of
GX339-4 is very hard, so the effects of pileup are instead most
noticeable at high energies so this default bandpass is not very
sensitive to the extent of pileup.  These ratios are clearly
incompatible with unity in the 6-10~keV band. Pileup strongly affects
the spectrum above $\sim $~6~keV even excluding the central 18 arcmin
and extracting only single events.

\begin{figure} 
\begin{center} 
\leavevmode  \epsfxsize=8cm \epsfbox{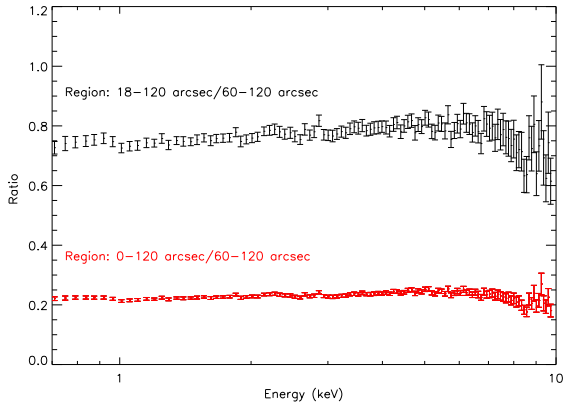} 
\end{center} 
\caption{A model independent way to show the effect of pileup on the
spectra. The ancillary response file (arf) for each spectrum is used to
correct the counts in each channel for the effective area. We ratio
these corrected counts for 0-120s/60-120s and 18-120s/60-120s. If pileup
were not important, these ratios should be unity at all energies. 
This is plainly not the case. The 0-120s/60-120s shows that many counts
are lost from the 0-120s spectrum due to pileup, and that this is not
a constant factor but depends on energy. While many fewer counts are
lost from the 18-120s spectrum, the energy dependence is clearly not
flat but results in a broad curvature of the spectrum with obvious
implications for modelling the red wing of the iron line.
}
\label{fig:arfs} 
\end{figure}

We exclude progressively larger central regions and find that we need
to go beyond 60 arcsec before the ratio of observed to expected
fraction in {\tt epatplot} is consistent (at $3 \sigma$) with unity in
the 6-10~keV for the single events.  The distribution for the doubles
still appears inconsistent, but the much poorer statistics means that
this is not significantly different from unity. Thus the spectra
extracted by R08 (singles, doubles and quadruple events, excluding a
50 arcsec core) should be mainly free from pileup in singles alone,
but doubles and quadruples could still be affected. 

This is at first sight surprising as the count rate per pixel is
nominally below the recommended on axis surface brightness limit of
$4.9\times 10^{-4}$~c/s/arcsec$^2$ (i.e. 0.35 c/s in a 15'' radius
circle). This shows that the effects of pileup can be subtle, and in
fact the double events still deviate from the expected curve above
6~keV even for data extracted from 90-120'', though again the
statistics are so poor that this is not significantly different from
unity (see Fig. \ref{fig:epatplot}d).

We use the spectrum extracted from single events in a 60-120 arcsec
annular region (hereafter 60-120s) as our reference spectrum to try to
assess the effect of pileup in the spectra extracted from singles in
the 18-120 arcsec region (hereafter 18-120s) and singles extracted in
the 0-120 arcsec region (hereafter 0-120s). We generate ancillary
response files (arf's) using the SAS task {\tt arfgen} for the three
spectral regions, and for each spectrum we divide each energy channel
count rate by its effective area at that energy as given by the arf.
This corrects for the differing effective areas in each extraction
region, so we can take a ratio of the spectra to show the effect of
pileup in a model independent manner. Fig. \ref{fig:arfs}
shows this for 0-120s (lower, red) and the 18-120s data (upper,
black). These ratios should be unity if the arf generation is correct
and no counts are lost in pileup. Instead, the full region gives a
mean value of 0.2, suggesting that the intrinsic count rate was close
to $\sim 120$ counts/sec, as predicted! More importantly, the ratio is
not flat, showing the derived spectra are distorted. This curvature is
not large but is very significant as the error bars are
small. Conversely, the 18-120s data are much closer to the expected
normalisation, showing that many fewer counts are lost, but there is
now clear curvature from the deficit of high energy single events
noted by {\tt epatplot} above 6~keV. This results in a broad residual
which could clearly affect the determination of the iron line width.

\begin{figure} 
\begin{center} 
\leavevmode  \epsfxsize=8cm \epsfbox{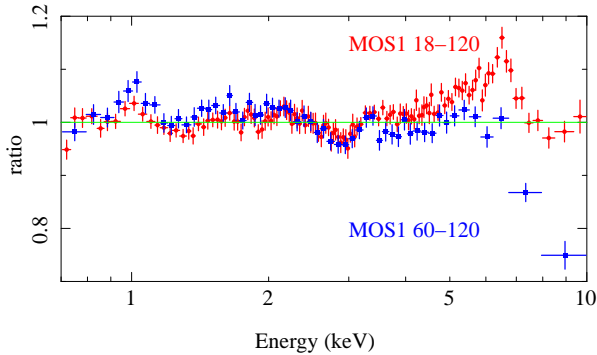} 
\end{center} 
\caption{Ratio of the MOS1 18-120s spectrum as in M06 (red) to the 
best fit continuum model excluding the 4-7~keV region. The blue points
show the 60-120s data ratio'd to the {\em same} model. The 18-120s
spectrum has more counts at high energies, as expected from pileup
(see Fig.~\ref{fig:epatplot}b), but it also has a much stronger 'red wing'. The 60-120s
data have almost no line emission, but instead are dominated by a
strong drop at 7~keV (see also R08).}
\label{fig:ratio_spec} 
\end{figure}

We test this explicitly by fitting the spectra directly. While
background is negligible for the 18-120s spectrum, it is much more
important for the 60-120s spectrum. We follow M06 and R08 and extract
background from a 60 arcsec region towards the corner of the chip. We
fit this in the same way as M06, i.e. with an absorbed ({\tt tbabs})
multicolour disc ({\tt diskbb}) plus power law to describe the
continuum over 0.7-10~keV, excluding the 4-7~keV iron line bandpass.
Fig. ~\ref{fig:ratio_spec} (red points) shows the ratio of the model
to the data over the full bandpass, with a strongly skewed line
residual as in M06. The blue points shows the 60-120'' data ratio to
the same continuum model. The residuals in the 4-7~keV region are
quite different, with no apparent line detected, but with a strong
drop above 7~keV.

The decrease in strength of the red wing, and increasing strength of
the drop above 7~keV is also shown by R08 for progressively larger
extraction regions. These data show that for a continuum model
set by the 18-120s spectrum, the red wing progressively disappears and
the 7-10~keV continuum drops. This is a clear sign of pileup in the
spectra with small extraction regions, as confirmed by the {\tt
epatplot} results (Fig.\ref{fig:epatplot}b). However, allowing the
model to float between the different datasets gives a steeper
underlying continuua spectra for the larger extraction regions, which
recovers the broad wing (Fig 1b in R08). This constant residual
profile was used by R08 to argue that the 18-120s spectrum as in M06
was not affected by pileup. 

However, this line profile in the 60-120s data is strongly dependent on the 
background used. The corners of the chip are furthest from the source position, 
but the source is so bright that the point spread function puts significant 
counts into these regions. This is seen both from comparing the putative 
background (chip corner) count rate with that seen at 
larger off axis angles from the other MOS1 chips and by comparing
to another observation of GX339-4 where the source was
ultrafaint (0085680501). Both these give a significantly lower
countrate, showing that the the chip corners are contaminated by the source. 
Photons scattered into the far wings of the PSF
are harder than average, so this results in the background being
contaminated by a harder version of the source spectrum. Background
subtracting with these data then artificially steepens the derived
source spectrum at high energies, which in turn leads to a broad
residual in the 4-7~keV from fits which exclude this energy band.

In summary, the MOS data extracted from 18-120 arcsec are clearly
piled up, even in single events. Extraction regions of $\sim$~60
arcsec are required before the single event spectra are usable, though
the double (and higher) events still show signs of pileup. Such large
exclusion regions means that the statistics are poor, leading to
increased systematic uncertainties from background (and response)
issues. This combination of factors mean the MOS data and background
are effectively unusable as shown.

\subsection{PN data}

\subsubsection{Data Selection}

The PN are in timing mode where only one CCD chip is operated. The
data are collapsed into a one-dimensional row (4\farcm 4) and read out
at high speed, with the second spatial dimension being replaced by
timing information. This allows a time resolution of 30~$\mu$s, and
photon pile-up occurs only for count rates $\ga$800~counts/s. Before
examining the pile-up effects with the {\tt epatplot} output, we used
the SAS task {\tt epfast} on the EPIC pn event files to correct for a
Charge Transfer Inefficiency (CTI) effect which has been seen in EPIC
pn Timing mode when high count rates are present \footnote[1]{More
information about the CTI correction can be found in the {\it EPIC
status of calibration and data analysis} and in the Current
Calibration File (CCF) release note {\it Rate-dependent CTI correction
for EPIC-pn Timing Modes}, by Guainazzi et al. (2008), at
http:$\slash\slash$xmm.esac.esa.int$\slash$external$\slash$xmm$\_$calibration}.

WebPIMMS gives an estimated total count rate of 260 c/s for the PN
medium filter in timing mode for pattern 0 events. This is lower than
the 800 c/s limit of this mode, but the source is hard so there is the
possibility of print through of X-ray photons during the offset map
calculation, which effectively reduces this limit by a factor of 2
(see Section 3.3.2 of the XMM-Newton users handbook). However, the
source is very variable so it can exceed this limit. This variability
can also cause the telemetry limit to be exceeded (at $\approx 450$
c/s for this mode if both MOS cameras are operated), at which point
the science data are lost.  Thus the telemetry limit is actually the
more stringent constraint, and effectively excludes the most pileup
data (see Section 3.3.2 of the XMM-Newton handbook), though short time
periods of higher count rate can be registered before the buffer
fills.

\begin{figure} 
\begin{center} 
\leavevmode  \epsfxsize=8cm \epsfbox{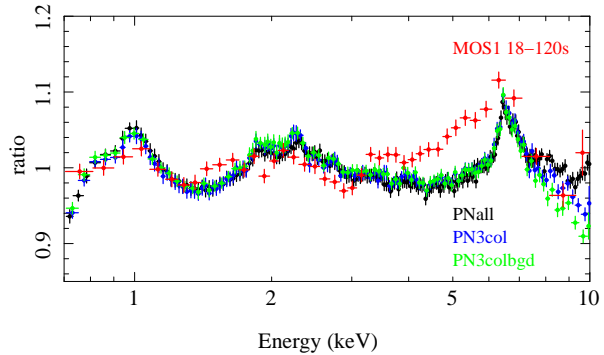} 
\end{center} 
\caption{Ratio of PNall (black) to its best fit continuum model,
excluding the 4-7~keV region. The blue and green points show the same
model ratio'd to the PN3col and PN3colbgd spectra.  All these PN
spectra are very similar except at the highest energies, where
PN3colbgd (green) is steeper than PN3col (blue), which is steeper than
PNall (black). We also show the MOS1 18-120s spectrum ratio'd to its
own best fit continuum model as in Fig.~\ref{fig:ratio_spec}. The line profile is
significantly broader in the piled up MOS data than in any of the PN
spectra.}
\label{fig:ratio_pn_mos} 
\end{figure}

\begin{figure} 
\begin{center} 
\leavevmode  \epsfxsize=8cm \epsfbox{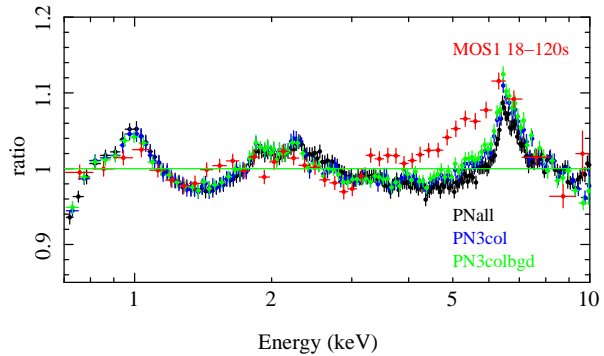} 
\end{center} 
\caption{As in Fig.~\ref{fig:ratio_pn_mos} except that all PN spectra are fit
with their own best fit continuum models. The difference at high
energies is therefore supressed by the power law becoming
progressively steeper from PNall (1.53) to PN3col (1.56) to PN3colbgd
(1.59). This results in progressively broader line residuals, but none
are as broad as seen in the (pileup) MOS spectrum  18-120s.}
\label{fig:pnfree} 
\end{figure}

First we extract a spectrum from columns 31--45.
While single events should be less affected by pileup, there
is a difficulty in uniquely registering such events in timing mode
as these include photons which hit two adjacent pixels (i.e. doubles) 
separated in the readout direction. Hence we extract single+double events 
and refer to this as PNall.  We
fit these data to the simple absorbed disc plus power law model for
the continuum, excluding the 4-7~keV region. The ratio of the data to
this fit is shown in as the black points in Fig~\ref{fig:ratio_pn_mos}. The
red points show the corresponding ratio for the 18-120s MOS spectrum
(to its own best fit continuum model). The difference in the line
shape is immediately apparent. 

However, Fig~\ref{fig:ratio_pn_mos} also shows that there are 
other residuals at energies of $\sim$~1.0, 1.8 and 2.2~keV. The $\sim
1$~keV feature is often seen in other LMXRBs though its exact nature
is still unclear (e.g. Sidoli et al 2001, Boirin \& Parmar 2003;
Boirin et al 2004; 2005; Diaz Trigo et al 2006; 2009). The features at
1.8-2.2~keV probably are artifacts of incomplete CTI correction in the
PN timing mode.

We check for pileup in the PN data using {\tt epatplot}. This shows
small deviations between the observed and predicted 'single' and
double event ratios for PNall, with larger features around the
instrument edges at 0.5~keV, 1.8~keV and 2.2~keV.  Thus these are
most probably associated with incomplete CTI correction. However, the
number of singles is systematically below the predicted curve at all
energies, while the doubles are systematically above. This could be
indicative of pileup, but the constant ratio of the effect at all
energies appears more likely to indicate a systematic error in
modelling the number of singles and doubles in this mode (which is
complicated as the exact pattern distribution depends on the
asymmetrical shape of the charge cloud, and the projection of this on
the readout direction: Kendziorra et al 2004).

We can check this by intensity slicing the PNall data, accumulating
spectra from 0-200, 200-400, 400-600 and $\ge 600$ c/s. If pileup is
important then this will progressively harden the brighter spectra.
However, simple disc plus power law fits indicate that the spectrum is
systematically {\em softer} for brighter spectra ($\Gamma=1.54$
compared to $\Gamma=1.62$ for the dimmest and brightest spectra,
respectively). This shows that the intrinsic spectral variability
(steeper when brighter, as in generally seen in the low/hard state
e.g. Churazov et al 1999; DGK07) is more important than any residual
pileup in these data.

Nontheless, we also re--extracted the data excluding the central 3
columns of the data in a technique analogous to excluding the central
region of imaging data (PN3col). The {\tt epatplot} results again have
features at the instrument edges, but now the difference at other
energies is much smaller. The blue points in Fig.~\ref{fig:ratio_pn_mos}
shows the ratio of these data to the same best fit continuum model as
used for the PNall data. They are extremely similar except at the
highest energies, where PN3col is steeper than PNall.  This difference
cannot be due to background as this would be more important for the lower
count rate PN3col spectrum, and would effectively {\em harden} rather
than soften its high energy spectrum.

It is possible then that pileup is causing the slight softening of the
PN3col spectrum compared to PNall. However, it is also possible that this
simply represents the limit on our current knowledge of the reconstructed
pattern distribution and energy dependence of the point spread function in
this mode. Hence we use both these spectra to illustrate the range of
models allowed by the data.

\begin{figure} 
\begin{center} 
\leavevmode  \epsfxsize=8cm \epsfbox{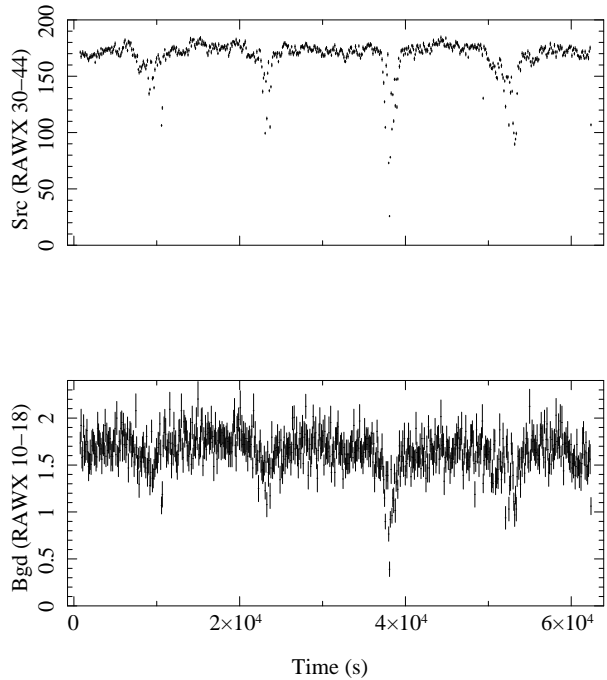} 
\end{center} 
\caption{The upper panel shows the PN timing mode lightcurve of the
dipping source XB1254-690 extracted from columns 30-44, while the
lower panel shows that from 'background' channels 10-18. It is clear
that a source of this brightness (around 250 c/s) contaminates 
the chip even at large off axis angles.}
\label{fig:lc} 
\end{figure}

For the PN3col data, the lower count rate means background may become
important. However, there are no source-free background regions on the
PN chip as the source contaminates the background event even in the
lowest/highest chip columns. This background contamination is worse
than in the MOS cameras due to the smaller field of view ($13.6\times
4.4$ arcmin in the PN versus $10.9\times 10.9$ arcmin for the MOS, but
the collapsed column for PN timing mode is the larger one, so the
maximum angle for background extraction is 2 arcmin compared to 5
arcmin for the MOS). The amount of background contamination is more
easily seen in the PN timing data from XB 1254-690. This is a dipper,
so the source drops dramatically in intensity (e.g. Diaz Trigo et al
2006). Fig.~\ref{fig:lc} shows the source lightcurve together with the
'background' lightcurve taken from columns 10-18 (as used 
by Wilkinson \& Uttley 2009, hereafter WU09). The background
shows the same dip structure as the source, showing that it is
dominated by source photons scattered out into the far wings of the
PSF. Again, since the PSF is energy dependent, these scattered photons
are harder than the real source spectrum so this source contaminated
background spectrum is {\em not} equivalent to subtracting a small
fraction of the source spectrum plus background.  Restricting the
background to columns 3-14 reduces the effect slightly (the source
counts increase from 95.8 per cent to 96.8 per cent), but the
lightcurve of these columns from XB 1254-690 still shows that they are
contaminated. A better estimate of the true background level can be
made by extracting timing data from the observation in which GX339-4
was so faint as to be undetectable (0085680501). Using these data as
background increase the source contribution to 99.8 per cent, showing
that neglecting background in the PN3col spectrum is more appropriate.
Nontheless, we also include a background subtracted PN3col spectrum
(hereafter called PN3colbgd), using background from columns 10-18 in
order to compare directly with the results of WU09).

We add 2 per cent systematics to account for uncertainties in the
response, though since the data are grouped by a factor 4 then this
results in a 1 per cent systematic error on the final spectrum and 
fit the data in the 0.7-10~keV range.

\subsubsection{Iron line profile}

\begin{table*}

\begin{tabular}{cccccccccccc}
  \hline
model & $R_{in}$  & $\beta$, inc  & $kT_{disk}$ & $N_{disk}$ & $\Gamma$ &
$N_{pl}$ & $E_{line}, \xi$ &   $N_{line},  N_{ref}$ & $\chi^2/\nu$ \\
    & $R_g$ & & keV & & & & keV & $\times
10^{-3}, \times 10^{-5}$
& \\
  \hline
& & & & PNall\\
\hline
laor    & $177_{-70}^{+3}$ & 
$3$, $60^\circ$ & $0.29(1)$ & $1920_{-380}^{+530}$  
& $1.52(1)$ & $0.186(2)$
& $6.63_{-0.02}^{+0.03}$ & 
$0.60(4)$ & $938/221$\\
& $25_{-6}^{+1}$ &
$10_{-3.3}^{peg}$,  $16\pm 1$   & $0.29(1)$ & $2100_{-550}^{+390}$ 
& $1.53(1)$ & $0.187(2)$ 
& $6.95\pm 0.03$ & 
$0.64(4)$ & $925/219$\\
reflionx & $170\pm 70$ & 
 $3$, $60^\circ$ & $0.33(1)$ & $790_{-20}^{+110}$ &
$1.53(1)$ & $0.178(1)$
& $370_{-20}^{+10}$ & 
$0.56(2)$ & $502/220$\\
& $25_{-6}^{+7}$ & 
$3$, $27(3)$ & $0.34(2)$ & $630_{-170}^{+170}$
& $1.53(1)$ & $0.176(2)$
& $490_{-40}^{+10}$ & 
$0.44(4)$ & $468/219$\\
\hline
& & & & PN3col\\
\hline
laor  & $19\pm 4$  & 
$3$, $60^\circ$ & $0.26(1)$ & $3890_{-930}^{+1000}$ 
& $1.57(1)$ & $0.198(2)$ & $6.4_{peg}^{+0.03}$ & 
$1.34(12)$ & $762/221$ \\
& $8.2_{-1.0}^{+1.2}$ & 
$3.4(2)$,  $17_{-2}^{+3}$ &  $0.26(1)$ & $4400_{-960}^{+1479}$ 
& $1.57(1)$ & $0.198(2)$ 
& $7_{-0.02}^{peg}$ & 
$1.11(10)$ & $716/219$\\
reflionx & $100_{-40}^{+60}$ &
$3$, $60^\circ$  & $0.26(2)$ & $3070_{-1150}^{+940}$
& $1.58(2)$ & $0.188(2)$ 
& $260_{-20}^{+40}$ & 
$1.16(28)$ & $488/220$\\
& $11_{-3}^{+2}$ & 
$3$, $25(3)$ & $0.30(2)$ & $1060_{-350}^{+650}$
& $1.56(1)$ & $0.182(3)$
& $500_{-60}^{+20}$ & 
$0.54(10)$ & $396/219$\\
\hline
& & & & PN3colbgd\\
\hline
laor &  $24_{-4}^{+1}$ & 
$3$, $60^\circ$ &$0.23(1)$ & $8560_{-1950}^{+2520}$ 
& $1.60(1)$ & $0.200(3)$ 
& $6.4^{+0.01}$ & 
$1.33(9)$ & $667/221$\\
& $6.33_{-0.08}^{+1.04}$  &
$3.4(2)$, $22_{-3}^{+1}$ & $0.23(1)$ & $7960_{-1960}^{+3300}$ 
& $1.60(1)$ & $0.200(3)$ 
& $7_{-0.02}^{peg}$ 
& $1.43(12)$ & $614/219$\\
reflionx & $60^{+40}_{-20}$ & 
$3$, $60^\circ$ &  $0.22(1)$ & $8000_{-2500}^{+4400}$
& $1.61(1)$ & $0.189(2)$
& $250^{+10}_{-10}$ & 
$1.19(26)$ & $456/220$\\
& $10_{-2}^{+2}$ & 
$3$  $27(3)$ & $0.24(2)$ & $4240_{-1820}^{+3210}$ 
& $1.59(1)$ & $0.179(3)$ 
& $530_{-30}^{+30}$ & 
$0.46(6)$ & $360/219$\\
\hline
& & & & MOS 18-120s\\
\hline
laor & $47\pm 15$ & 
$3$, $60^\circ$  & $0.29(2)$ & $2790_{-800}^{+1100}$ & 
$1.55(2)$ & $0.198(4)$ 
& $6.4^{+0.01}_{peg}$ & 
$1.06(20)$ & $309/218$\\
& $2.6_{-0.6}^{+0.3}$ & 
$3.2(2)$, $0^{+30}_{peg}$ & $0.31(2)$ &  $1770_{-370}^{+1030}$  
& $1.57(2)$ & $0.200(5)$ 
& $6.95_{-0.25}^{+0.05}$ & 
$3.27(27)$ & $220/216$\\
reflionx & $44_{-20}^{+20}$ & 
$3$, $60^\circ$ & $0.27(2)$ & $3850_{-1380}^{+1650}$
& $1.61(2)$ & $0.211(7)$ 
& $10_{peg}^{+3}$ & 
$53(20)$ & $297/217$\\
& $2.0_{-0.77peg}^{+0.6}$ & $3$, $21(6)$ & $0.32(1)$ & $1400_{-400}^{+400}$ 
& $1.56(2)$ & $0.172(10)$
& $500_{-180}^{+80}$ & 
$1.07(13)$ & $200/216$\\
\hline
\end{tabular}
\caption{The various PN spectra fit with 
a simple {\sc diskbb+po+laor} model (first two lines) and a more
physical {\sc diskbb+po+kdblur*reflionx+gau} model (second two
lines). For each model the 
emissivity and inclination are first fixed at $3$ and
$60^\circ$, respectively. Both these parameters are then allowed to be
free for the {\sc laor} model, but the resulting emissivity generally
close to $3$ so this is fixed in the more complex reflection fits when
the inclination is a free parameter. The corresponding column density is
$0.42-0.44$, $0.45-0.48$, $0.48-0.51$ and $0.49-0.54\times
10^{22}$~cm$^{-2}$ for each of the 4 spectra. A narrow
(width fixed at 0.01~keV) gaussian line at 6.4~keV is also
included in the reflection fit, as there is a significant residual
here, although its equivalent width is small. 
}
\end{table*}

We redo the  Fig~\ref{fig:ratio_pn_mos}, but this time ratio each PN
spectrum with its own best fitting model, where only the absorption is
tied across the three datasets. The differences in the
PN spectra at high energies means that this gives a progressively
steeper power law spectra for PNall, PN3col and PN3colbgd, (1.53, 1.56
1.59 respectively). This results in progressively slightly
broader line residuals for the 4.0-7.0~keV region excluded from the
continuum fit as shown in Fig. \ref{fig:pnfree}. However, these are
all still obviously narrower than the line from the MOS 18-120s data. 

We quantify the differences in line profile between the MOS and PN by
fitting each spectrum separately with a model of {\sc
tbabs*(diskbb+po+laor)}, where the intrinsic line energy is
constrained to be between $6.4-7.0$~keV.
We first fix the inclination
at $60^\circ$ (a reasonable value for the binary inclination in
GX339-4: Kolehmainan \& Done 2010) and the {\sc laor} line emissivity
at $3$. Then we allow both inclination and emissivity to be free
parameters for each spectrum. All the PN spectra show a better fit for
a solution with much lower inclination as the line profile in the PN
data is quite symmetric rather than being skewed to the red. The
smaller Doppler shifts from a lower inclination then require a smaller
radius to get the same line width, but all of the PN data are
consistent with a disc which does not extend below $6R_g$. By
contrast, the piled up MOS 18-120s spectrum strongly requires a much
smaller inner radius.

To have confidence in the line parameters requires modeling the
reflected continuum as well as the line.  We use the {\tt reflionx}
model for an ionised slab of constant density (Ross \& Fabian
2005). This includes Compton scattering within the disc itself which
can be important at high ionisation (Ross, Fabian \& Young 1999),
making the line intrinsically broad.  We convolve this with
relativistic effects using {\sc kdblur} (based on the {\sc laor} line
kernel) model, and also include a narrow, neutral iron line, as this
is plainly present as a small residual. Again we first fix the
inclination and emissivity at $60^\circ$ and $3$, respectively.  We
then allow the inclination to be free, but keep the emissivity fixed
at $3$ since the {\sc laor} models showed that this is generally close
to the derived value. We checked that there is no significantly better
fit produced by allowing this to be free. Full details of all the
parameters for these fits are given in Table 1. 

Unsurprisingly, the full reflection model always gives a substantially
better fit than the equivalent {\sc laor} line fits. The data still
prefer a lower inclination solution ($\sim 30^\circ$) than seems
likely from the binary orbit (Kohlemainen \& Done 2010), but we
caution that this depends on the model used for both the continuum and
reflected emission (the continuum is unlikely to be a single power
law: see e.g. DGK07 section 4.3, and the detailed shape of the
reflected spectrum depends on the assumed vertical structure of the
disc: Malzac et al 2005). We will explore this further in a subsequent
paper. Here we only concentrate on the iron line, and it is clear that
the fits do not require an extreme relativistic profile for any of the
PN spectra, with the inner disc always being larger than $6R_g$ even
for the low inclination solutions, while the physically more likely
higher inclination solutions strongly require a truncated disc.

The normalisation of the disc component is also a tracer of the inner
radius of the accretion disc. This is consistent with a constant at
$\sim 4700$ in the disc dominated high/soft state of GX339-4
(Kohelmainen \& Done 2010; DGK07 fig. 13b). M06 claim that the disc
component seen in the MOS 18-120s spectrum is consistent with the same
inner radius, arguing against disc truncation. However, at face value,
the disc normalisations for both MOS and PN are substantially lower
(see also DGK07 fig 13b), except for the PN3colbgd which clearly has
oversubtracted background (see Fig.~\ref{fig:lc}).  However, the disc
normalisation is again strongly dependent on the assumed continuum and
reflected emission (see e.g. DGK07 section 4.3), as well as being
affected by irradiation (Gierlinski, Done \& Page 2008) and Compton
scattering (Makishima et al 2008). Again, we will consider this in
more detail in a subsequent paper, as here we focus on the iron line profile.

\section{Fits including RXTE data}

The telemetry limitation on the PN data means that the normalisation
should be lower in the PN than in RXTE since the brightest data are
excluded. However, the correlation between spectral index and
brightness in these data (see previous section) means that the lost
data are preferentially the softest. Thus the PN should have a
slightly harder spectral index than the RXTE data. Formally, the
two cannot be fit simultaneously without re-extracting the RXTE data
to exclude the same highest intensity time intervals as are lost in
the PN. Nonetheless, we use all the RXTE data in order to compare with
previous work, but caution that this may make small differences
in the results. 

\subsection{PN-PCA-HEXTE: Miller et al (2006)}

\begin{figure*}
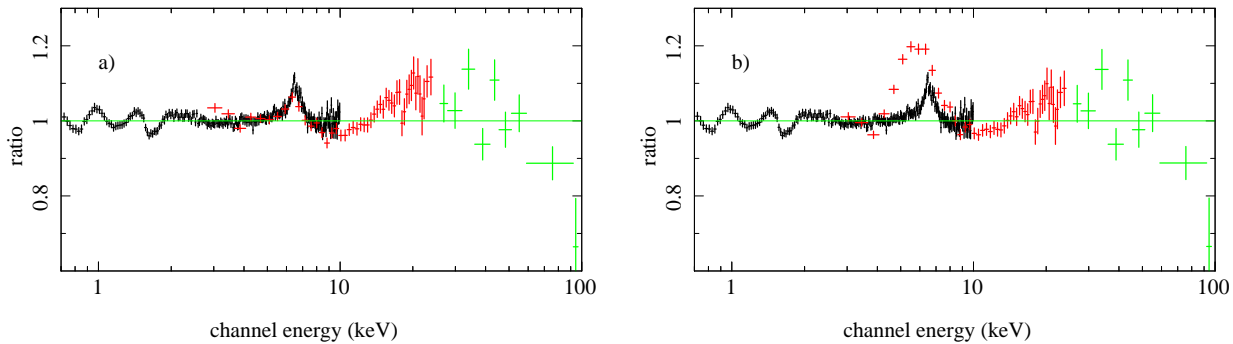

\begin{center}
\begin{tabular}{cc}
\leavevmode  \epsfxsize=8cm \epsfbox{rxte_pn_rat_a.ps}
&
\leavevmode  \epsfxsize=8cm \epsfbox{rxte_pn_edge_rat_b.ps}
\end{tabular}
\end{center}
\caption{Residuals from simultaneous fits of the PN3col and RXTE PCA and HEXTE
  spectra to an absorbed power law plus disc model, excluding the
  4-7~keV band. a) uses the default RXTE PCA
  response. b) shows the effect of including
  an edge at 4.78~keV with optical depth of 0.1 to account for
  calibration as in M06. Plainly this has a strong effect on the
  inferred PCA line profile. Without the edge, the PCA line is
  consistent with the PN but not with the piled up MOS. Conversely,
  with the edge the PCA line is consistent with the piled up MOS, but
  not with the PN. } 
\label{fig:pn_rxte}
\end{figure*}

The narrower PN line is at first sight in conflict with the
consistency between 18-120s MOS data and the RXTE results. Hence we
re--extract the RXTE data following M06 i.e. we use the RXTE observation
ID 90118-01-06-00, extract data from all layers from PCU2, and add
a systematic uncertainty of 0.6 per cent. We also extract the HEXTE data
from HXT0 for the same observation, and use the same energy ranges as
M06, i.e. 2.8-25.0~keV for the PCA, and 20-100~keV for HEXTE.

We illustrate our results with PN3col (the middle of the
three PN spectra in terms of high energy slope). We co--fit these
and the RXTE data with an absorbed disc plus power law model,
excluding the 4-7~keV region. We tie all spectral parameters across
all 3 instruments, but allow a variable normalisation factor. The
ratio of the data to this continuum model is shown in
Fig.~\ref{fig:pn_rxte}a. The iron line shape is consistent between the
PN and PCA data.

M06 introduce an edge at 4.78~keV, with depth $\tau=0.1$, to account
for residuals sometimes seen in RXTE data at the instrumental Xe L
edge. We include this edge and repeat the fit above, and obtain the
residuals shown in Fig.~\ref{fig:pn_rxte}b. It is clear that this has
a strong impact on the derived line shape in the PCA, and on the
apparent agreement with the piled up MOS data. Instead, 
the standard PCA response gives consistent fits with the PN spectrum.

\subsection{PN-PCA: Wilkinson \& Uttley 2009}

We also check for consistency with WU09.  They used the same PN timing
data as examined here, yet their best fit model line extended down to
$4R_g$ for an inclination of $40^\circ$. This was derived from
approximating the ionised reflection spectrum by a {\tt laor} line and
neutral reflection continuum ({\sc hrefl}, which does not include the
reprocessed emission), and fit to the PN data
extracted by excluding the central 3 columns, and background
subtracting. Thus these are similar to our PN3colbgd data so we fit
this spectrum together with the RXTE PCA data to reproduce the WU09
results.

We find very similar model parameters, including the broad line
$R_{in}=2.5\pm 0.5~R_g$, large amount of reflection,
$\Omega/2\pi=0.95\pm 0.15$ and steep continuum $1.75\pm 0.02$. This
fit is formally unacceptable at $\chi^2_\nu=677/267$, but this is
mainly due to the residuals noted earlier in all the PN
spectra. However, it describes the overall curvature fairly well,
which was all that was required in WU09 for their investigation of the
variability properties.

We replace the {\it ad hoc} line plus reflection description with the
{\sc reflionx} model of an ionised reflector, with relativistic
smearing by convolution with {\tt kdblur} as above. We keep the
inclination set to $40^\circ$ and find a much better fit
$\chi^2=544/268$ (one fewer degrees of freedom as the line energy and
intensity are set by the disc ionisation parameter). This gives a much
larger inner disc radius, as the much softer shape of the reprocessed 
emission from ionised material twists the spectral fit so that the
inferred spectral slope is much harder at $1.6$ (and $\Omega/2\pi$
much smaller), so the inferred red wing is much less prominent. This
is compounded by the Compton broadening which is already present in
the line from ionised material, so that the same data give
$R_{in}=44~R_g$.

\section{Conclusions}

The detection of an extremely broad iron line in GX339-4 in M06 is an
artifact of pileup in the XMM-Newton MOS data despite their efforts to
mitigate the effects of this by excising the central 18'' core of the
image (M06, R08). We show that the MOS data are only free from the
effects of pileup when such a large central core is excluded that the
spectra are unusable due to systematic uncertainties in background
subtraction.

The simultaneous PN data in Timing mode are much less piled up. 
They may still be slightly affected at the highest countrates from this
variable source, though these data are mainly lost through telemetry
limitations.  We discuss the range of PN spectra that can be derived
from these data. If there is no pileup then the full point spread
function can be used, but we also excise the core in case pileup is
present. We discuss whether background should be subtracted from
these. Plainly background will have more effect on the lower count
rate spectrum from the excised core data than from the full
PSF. However, we show that the 'background' is strongly contaminated
by the source, and source photons scattered so far in the wings of the
PSF are preferentially hard. Thus subtracting this 'background'
artificially steepens the spectrum at high energies.

We consider the PN spectra from the full PSF and with the core excised
as representing the ranges of solutions alllowed by the data, but also
fit the 'background' subtracted PN data in order to compare with
WU09. We show that all of these give a line profile which is
significantly narrower than that of M06 from the MOS 18-120s data. All
the PN datasets give an inner radius which is larger than the last
stable orbit for a non-spinning black hole at $6R_g$ even for
(physically unlikely) low inclination solutions, and this radius
increases for (more probable) higher inclination angles for this
source.

This narrower line is consistent with the RXTE PCA data, and the
apparent match of the PCA with the much broader line in the MOS data
of M06 is a consequence of their inclusion of an edge at 4.78~keV to
account for potential deficiencies in the PCA response around the Xe L
edge. The PN and RXTE data can be fit simultaneously although the
telemetry limits on the PN and corelation of spectral index with
luminosity means that these should have subtly different spectra.
These fits give a small inner radius consistent with WU09
when fit with the same neutral reflection/ionised line
model. However, this radius increases substantially when the
fully self-consistent ionised reflection models are used.

Thus the PN data support the truncated disc interpretation of
the low/hard state data, in direct conflict with the piled up MOS data
which require that the disc extends down close to the last stable
orbit of a maximally spinning black hole. Similar issues with pileup
distorting the derived iron line shape have also been seen in
Suzaku data of GX339-4 in an intermediate state (Yamada et al
2009). We strongly urge caution in using piled up data for detailed
spectral analysis, and especially where those data are discrepant in
some way with previous results. 

\section*{Acknowledgements}

Based on observations obtained with XMM-Newton, an ESA science mission
with instruments and contributions directly funded by ESA member
states and the USA (NASA).  M. D{\'i}az Trigo thanks the EPIC
Calibration Scientist, Matteo Guainazzi, for helpful discussions
regarding pile-up.

\end{document}